\documentclass[twocolumn,showpacs,aps,prl]{revtex4}  
\usepackage{epsfig,amssymb}

\input psfig.sty
\begin{document}

\title{Vortex Lattice Dynamics in Rotating Spinor Bose-Einstein Condensates}
\author{V. Schweikhard, I. Coddington, P. Engels, S. Tung, and E.~A. Cornell\cite{qpdNIST}}
\affiliation{ JILA, National Institute of Standards and Technology
and University of Colorado, and Department of Physics, University
of Colorado, Boulder, Colorado 80309-0440}
\date{\today}

\begin{abstract}
We observe interlaced square vortex lattices in rotating
dilute-gas spinor Bose-Einstein condensates (BEC). After preparing
a hexagonal vortex lattice in a one-component BEC in an internal
atomic state $|1\rangle$, we coherently transfer a fraction of the
superfluid to a different state $|2\rangle$. The subsequent
evolution of this pseudo-spin-$1/2$ superfluid towards a state of
offset square lattices involves an intriguing interplay of
phase-separation and -mixing dynamics, both macroscopically and on
the length scale of the vortex cores, and a stage of vortex
turbulence. The stability of the square structure is proved by its
response to applied shear perturbations. An interference technique
shows the spatial offset between the two vortex lattices. Vortex
cores in either component are filled by fluid of the other
component, such that the spin-$1/2$ order parameter forms a
Skyrmion lattice.
\end{abstract}


\maketitle

Some of the most intriguing phenomena of superfluidity are
revealed when a superfluid is set rotating. Depending on the
complexity of the order parameter a superfluid can carry angular
momentum in different ways. A single-component superfluid exhibits
quantized vortices that, in a dilute-gas single-component BEC,
organize into a regular hexagonal Abrikosov lattice
\cite{Abrikosov}[see Fig. 1(a)], whose static properties and
long-wavelength excitations have been examined
\cite{Vortices,JILATkachenko,LLL,VortexCores}. Such lattices are
also formed in Type II superconductors in a magnetic field
\cite{Abrikosov} and in superfluid $^4$He \cite{Donelly}.
Multi-component superfluids, with their pseudo-spin order
parameters, carry angular momentum in spin textures (Skyrmions)
\cite{Skyrmions,Mueller}, which may again form lattices
\cite{MuellerHo,Kasamatsu,Reijnders}, such as in the superfluid
phases of liquid $^3$He \cite{Salomaa} and in certain regimes of
two-dimensional electron systems \cite{Brey}.
\par
In this work we study rotating two-component superfluids in a
magnetically trapped dilute-gas BEC \cite{TwoComponentIntro}
described by a pseudo-spin-$1/2$ order parameter. State
$|1\rangle$ represents spin-down and state $|2\rangle$ represents
spin-up, and the azimuthal component of the spin is fixed by the
relative phase $\Phi$ between the state $|1\rangle$ and
$|2\rangle$ wavefunctions. Both the large-scale density profiles
and the vortex lattice structure depend on the nature of inter-
and intracomponent interactions. The interaction energy,
determined by three scalar coupling constants $g_{11}$, $g_{22}$
and $g_{12}$ and the population densities $n_1$ and $n_2$, is
given by
\begin{equation}
E_{int}= \int d^3x[g_{11} n_1^2 + g_{22} n_2^2 + g_{12} n_1 n_2 ]
\end{equation}
For attractive intercomponent interaction ($g_{12} <0$) it is
energetically favorable if the spatial profiles of both components
coincide, and a hexagonal vortex lattice is expected, as in the
one-component case. For repulsive intercomponent interaction
($g_{12} >0$), realized in our $^{87}$Rb system, it is
energetically favorable for the two components to reduce spatial
overlap. In a static BEC this may be achieved by macroscopic phase
separation, whereas in rotating two-component BEC the presence of
the vortex lattice allows for more subtle separation effects, such
as interlacing the lattices. Vortices in either component may be
filled by the other component, and the pseudo-spin-$1/2$ order
parameter forms a Skyrmion lattice. Studies by Mueller and Ho
\cite{MuellerHo} and Kasamatsu et al. \cite{Kasamatsu} predict,
among other structures, interlaced square lattices over a wide
range of parameters. Presumably this is because in a square
lattice, unlike in a triangular lattice, each vortex in component
$|1\rangle$ can have all its nearest-neighbor vortices be in
component $|2\rangle$, and vice versa \cite{Lehnert}.

\begin{figure}
\begin{center}
\psfig{figure=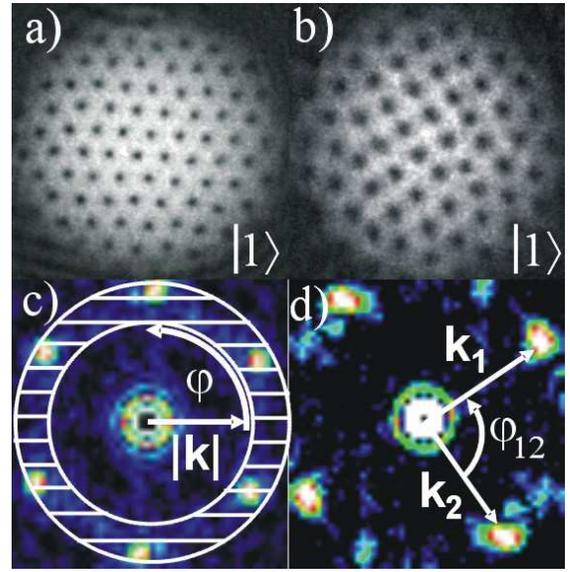,width=0.85\linewidth,clip=}
\end{center}
\caption {(a) Hexagonal vortex lattice in a one-component BEC. (b)
Square lattice, viewed in the $|1\rangle$ component of a
two-component BEC. (c) Reciprocal ($k-$) space of the hexagonal
lattice, obtained by 2D Fourier transform of (a), showing 6 peaks
spaced by $60^{\circ}$. (d) $k-$ space of the square lattice. The
reciprocal lattice vectors $k_1$ and $k_2$ enclose an angle
$\varphi_{12}=90^{\circ}$.} \label{HexSquare}
\end{figure}

\par
Under suitable conditions we indeed observe a spatial separation
of vortices in the two components, followed by the formation of an
ordered square lattice structure, as shown in Fig. 1(b). Its
formation and decay dynamics, as well as its static and dynamic
properties are examined in this work.
\par
Our $^{87}$Rb two-state system \cite{TwoComponentIntro} consists
of two magnetically trappable hyperfine-Zeeman levels of the
$^{87}$Rb atom - $|F=1,m_{F}=-1\rangle$, henceforth called
$|1\rangle$, and $|F=2,m_{F}=1\rangle$, henceforth called
$|2\rangle$. Spatial overlap of the two states is realized in a
harmonic, axially symmetric magnetic trap with oscillation
frequencies
$\{\omega_{\rho},\omega_{z}\}=2\pi\{7.7,4.9\}\,\rm{Hz}$. An
electromagnetic coupling between the two states can be achieved
via a two-photon transition, involving a microwave photon at
$\sim6.833\,\rm{GHz}$ and a radio frequency photon at $\sim 1
\,\rm{MHz}$. A short-pulse application of this coupling drive
yields a desired, spatially uniform amount of population transfer
between the two states, instantaneous with respect to the external
dynamics of the two states. In the pseudo-spin-$1/2$ picture the
effect of the coupling drive is to cause spin-rotations
\cite{TwoComponentIntro}. Inelastic atomic collisions limit the
lifetime of the $|2\rangle$ population to a few seconds. The decay
does not cause $|2\rangle$ atoms to convert back to the
$|1\rangle$ state but rather to leave the trap altogether.
\par
To study the rotational properties of this two-superfluid system,
we initially create regular hexagonal vortex lattices in BECs in
state $|1\rangle$, as described in earlier work
\cite{GiantVortex,LLL,VortexCores}. Here we start with near-pure
condensates containing $(3.5-4)\times 10^6$ atoms, rotating at a
rate $\Omega \approx 0.75\times\omega_{\rho}$ about the z-axis. A
``transfer pulse" of the coupling drive then transfers a fraction
of the population into state $|2\rangle$ , in this work
$80-85\,\%$ as discussed below. After a variable wait time we take
two images of the system. A nondestructive phase-contrast image is
taken either of one component alone or of both components
simultaneously \cite{TwoComponentIntro}, along an axis
perpendicular to the rotation axis (``side view") while the system
is still trapped. In order to optically resolve the vortex
structures we expand either the $|1\rangle$ or $|2\rangle$
component of the condensate by a factor of 9, to a diameter of
$\sim 600\,\rm{\mu m}$, before a second, destructive image is
taken along the rotation axis (``top view"). The other component
is removed at the beginning of the expansion. The details of
expansion and imaging have been described in Ref.
\cite{VortexCores}.
\par

\begin{figure}
\begin{center}
\psfig{figure=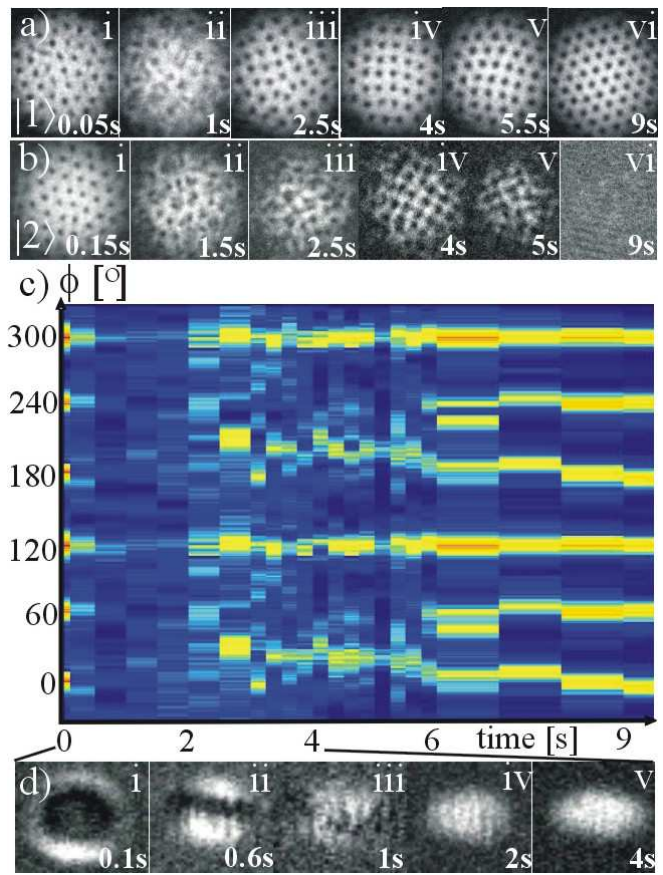,width=1\linewidth,clip=}
\end{center}
\caption {(a) Time sequence of images of state $|1\rangle$, after
$\sim80\%$ population transfer to $|2\rangle$, showing evolution
from a hexagonal lattice over a turbulent stage to a square
structure and back to a hexagonal lattice. (b) Also state
$|2\rangle$ forms a square lattice (iv) before its decay. (c)
Detailed time evolution in reciprocal space. Intensity in an
annulus along the $\varphi$ coordinate [defined in Fig. 1(c)] is
shown on the ordinate. The initial 6-peak structure of the
hexagonal lattice vanishes quickly because of turbulence. From
$3-5.5\,\rm{s}$ square lattices give rise to a 4-peak structure.
Around $6-9\,\rm{s}$ a transition back to hexagonal lattices
occurs. (d) Two-color side view images of the initial turbulent
evolution. State $|1\rangle$ ($|2\rangle$) appears bright (dark)
on gray background. The fine filament structures in (ii-v) are due
to mutual filling of vortex cores.} \label{TimeEvolutiuon}
\end{figure}

In Fig. 2 we analyze the formation and decay dynamics of the
square lattice structure. Figs. 2(a) and (b) show time sequences
of top-view images of the expanded $|1\rangle$ and $|2\rangle$
states, taken in different experimental runs. The time evolution
after the transfer pulse involves several stages. For the first
$\sim 0.1-0.25\,\rm{s}$ surprisingly little dynamics is visible
and certainly no structural transition in the vortex lattice is
seen in either component. From $\sim 0.25 - 2\,\rm{s}$ a turbulent
stage evolves in both components in which vortex visibility
degrades significantly, shown in Fig. 2[a(ii)] and Fig. 2[b(ii)].
As we will show, this turbulence is directly linked with the
transition from overlapping hexagonal vortex lattices to
interlaced square lattices. From $2 - 3\,\rm{s}$ square domains
emerge from the turbulent state, and defects propagate out of the
lattice \{Fig. 2[a(iii)]\}. From $3 - 5.5\,\rm{s}$ stable square
lattices are observed in both components \{Fig. 2[a(iv)] and Fig.
2[b(iv)]\}. At this stage, around $4\,\rm{s}$, despite the large
($80-85\,\%$) initial population transfer to state $|2\rangle$,
the number of $|2\rangle$ atoms has decreased to only $1.5 \times
10^5$, while state $|1\rangle$ contains $(5-7)\times 10^5$ atoms.
As the $|2\rangle$ state population continues to decay, the vortex
lattice planes bend \{Fig. 2[a(v)]\} and a transition back to a
hexagonal lattice in state $|1\rangle$ takes place \{Fig.
2[a(vi)]\}. During the transition from square lattices to a
hexagonal lattice no turbulence occurs.
\par
A more quantitative analysis of these dynamics is possible in
reciprocal space. Figure 2(c) shows the time evolution of the
intensity within an annulus in reciprocal space, defined in Fig. 1
(a), which contains the reciprocal lattice peaks. Initially six
peaks are visible, separated by $60^{\circ}$, forming the
reciprocal lattice of a hexagonal vortex lattice. Because of
turbulence these peaks vanish between $\sim200\,\rm{ms}$ and
$2\,\rm{s}$. After $2-3 \,\rm{s}$ a four-peak structure appears,
that is stable for a period of $\sim 2.5\,\rm{s}$. The observed
ratio of reciprocal lattice vector lengths [defined in Fig. 1(d)]
$k_1/k_2=0.98(2)$ and the angle $\varphi_{12}=95(3)^{\circ}$
between $k_1$ and $k_2$, clearly identify a square lattice. At
$\sim 5.5\,\rm{s}$ the appearance of two additional peaks signals
the onset of a transition back to the hexagonal state, which is
completed by $\sim 9\,\rm{s}$.
\par
To examine the origin of the initial turbulence, we show in Fig.
2(d) a time sequence of in-trap side-view images, where the
$|1\rangle$ ($|2\rangle$) state appears bright (dark). As visible
in Fig. 2[d(i)], a macroscopic component separation from the
initially homogeneous two-component superposition to a ball-shell
structure takes place within $50 - 100\,\rm{ms}\,$
\cite{TwoComponentIntro}. During this period of dramatic axial
separation, both the individual vortices and the overall vortex
lattice remain remarkably quiet, as viewed along the rotation axis
\{see Fig. 2[a(i)] and Fig. 2[b(i)]\}. Around $600\,\rm{ms}$, fine
filament structures appear at the intercomponent boundary \{Fig.
2[d(ii)]\}, coincident with the full development of vortex
turbulence seen in top-view images \{Fig. 2[a(ii)]\}. The filament
structures distort and fill out the whole BEC as the vortex
turbulence peaks at around $1\,\rm{s}$ \{Fig. 2[d(iii)]\}, and
straighten up at around $2\,\rm{s}$ \{Fig. 2[d(iv)]\} coincident
with the restoration of vortex visibility in top view images.
Subsequently the filaments become less visible as state
$|2\rangle$ decays \{Fig. 2[d(v)]\}. We interpret these structures
as vortex cores in either component being filled by fluid of the
other component, forming a Skyrmion lattice \cite{Mueller}.
Filled vortex cores grow in size above the resolution limit of our
side view images and become observable, while empty vortices in
single-component BECs in equilibrium are well below this
resolution limit and are not observed in-trap
\cite{Filledemptycores}. The initial {\it macroscopic},
predominantly axial phase separation has thus evolved into a {\it
microscopic} separation of two interlaced vortex lattices. It is
this {\it microscopic} separation which gives rise to the observed
turbulence.
\par

\begin{figure}
\begin{center}
\psfig{figure=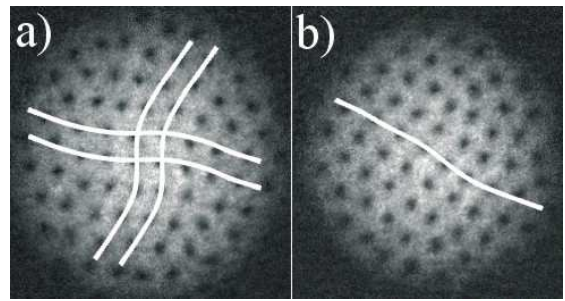,width=0.85\linewidth,clip=}
\end{center}
\caption{(a) Tkachenko excitation in the square lattice, observed
$250\,\rm{ms}$ after beginning of excitation, and (b) after
relaxation back to a square lattice structure (at
$850\,\rm{ms}$).} \label{Tkachenko}
\end{figure}

To study the stability of the square lattice, we excite shear
perturbations (Tkachenko modes) \cite{Anglin,BaymTkachenko} in the
square lattice by focusing a resonant laser beam onto the center
of the condensate \cite{JILATkachenko}. As shown in Fig. 3, the
perturbation relaxes to the equilibrium square configuration
within $500 - 800\,\rm{ms}$ after excitation. However, in contrast
to single-component triangular lattices \cite{JILATkachenko}, no
clear oscillation is observed. Two effects may contribute: The
excitation may be overdamped because of ill-defined boundary
conditions caused by imperfections in the outer region of the
square lattice, where the $|2\rangle$ state population has already
decayed. The oscillation may also be masked by random lattice
excitations of comparable amplitude that cannot be completely
removed in the short time between formation and decay of the
square lattice. However, the return of the lattice to its square
configuration is sufficient to demonstrate the stability of the
square structure in the two-component system.
\par

\begin{figure}
\begin{center}
\psfig{figure=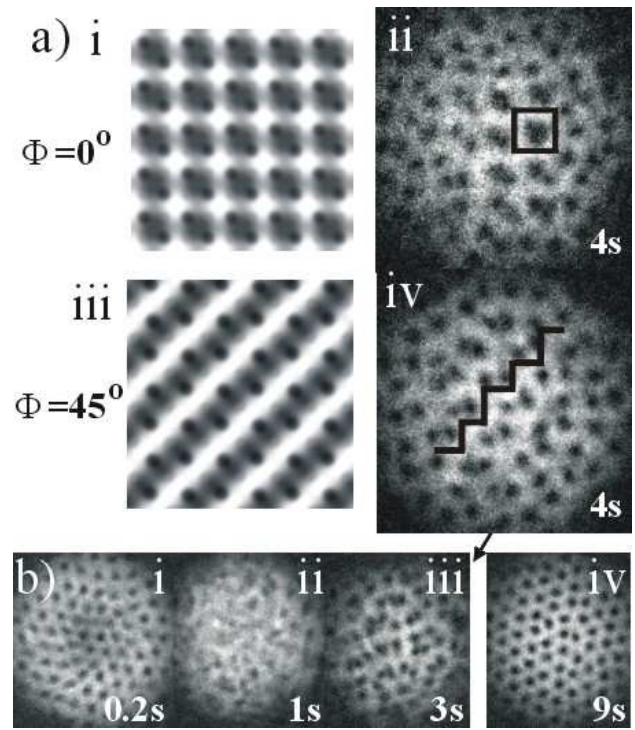,width=0.95\linewidth,clip=}
\end{center}
\caption {(a) Vortex lattice interference. (i) and (iii):
Simulations for different values of the relative phase $\Phi$
between the $|1\rangle$ and $|2\rangle$ state wavefunctions. (ii)
and (iv): Experimental results. (b) Time evolution of the
interference patterns shows no vortex offset before the turbulence
(i) but patterns characteristic of offset vortices afterwards
[b(iii)],[a(ii)],[a(iv)].} \label{Interference}
\end{figure}

So far we have presented evidence that both components separately
form regular and stable square lattices. Filament structures
indirectly indicate a microscopic-scale spatial separation of
vortices. In the following, we employ an interference technique
between the two superfluids to more concisely address two
questions: Are the vortex lattices really offset from each other?
Do the vortices really only separate from each other during the
turbulent stage, after a surprising delay of $200\,\rm{ms}$?
Figures 4[a(i)] and (iii) show results of a simple simulation of
interference between two square vortex lattice wavefunctions,
spatially offset by $1/2(\vec{a}+\vec{b})$, where $\vec{a}$,
$\vec{b}$ are the two basis vectors of the square lattice. Two
values ($\Phi=0$ and $\Phi=45^{\circ}$) of the relative quantum
phase $\Phi$ between the two wavefunctions
\cite{RelativePhaseFootnote} are considered. A population ratio of
$N_{|2>}:N_{|1>}=20:80$ was assumed, as in the experiment at
$4\,\rm{s}$. Evidently two very dissimilar patterns are seen,
consisting either ($\Phi\sim0$) of dark patches in between two
vortices, surrounded by a bright square structure, or
($\Phi\sim45^{\circ}$) of staircase-like vortex arrangements
separated by bright stripes. We checked that the qualitative
appearance of such simulations changes drastically when altering
the spatial offset from $1/2(\vec{a}+\vec{b})$.
\par
To observe such structures experimentally, we apply a $\pi/2$
``interference" pulse to the two-component system just before
expansion for the top view image. This pulse results in a
phase-coherent transfer of population between the two states, thus
creating interference. If the $\pi/2$ pulse is applied under
conditions when regular square lattices are expected in both
components, the observed images, Fig. 4[a(ii)] and (iv), agree
qualitatively with the simulations, clearly demonstrating a
spatial offset close to $1/2(\vec{a}+\vec{b})$ between the vortex
lattices.
\par
The qualitative agreement between experimental results and
simulation indicates that vortex lattice interference may be
employed to examine the separation process of the vortex lattices.
When varying the wait time between the transfer pulse and the
$\pi/2$ interference pulse, we observe the following qualitative
features [Fig. 4(b)]: Until turbulence occurs (after
$200-300\,\rm{ms}$) there are no indications for separation of the
vortex lattices. Figure 4.(b)(i) shows a hexagonal lattice with
good vortex contrast, very similar to images of each single
component at this time. This confirms that during the initial
$200-300\,\rm{ms}$ of dramatic macroscopic component separation,
vortices in the two components continue to form identical and
overlapped lattice structures, when viewed along the rotation
axis. Only after this surprisingly long delay, and after the
turbulent stage do interference patterns show a grouping
characteristic of offset vortices \{Fig. 4[b(iii)]\}. This
observation is further proof of the direct link between the
turbulent period and the microscopic separation of the vortex
lattices.
\par
In conclusion, we observe a new vortex lattice structure in
rotating two-component BEC. Each component carries a square vortex
lattice, and the lattices are interlaced. Vortices in both
components are filled by fluid of the other component, forming a
Skyrmion lattice. This structure is stable, as evidenced by
relaxation of applied shear excitations back to the square
structure. An intriguing turbulent stage accompanies the
transition from overlapped hexagonal lattices to interlaced square
lattices. Vortex lattice interference is used to study the offset
dynamics of the two lattices.
\par
This work was funded by NSF and NIST. We acknowledge useful
conversations with Nick Read.




\begin{thebibliography}{21}
\bibitem{RelativePhaseFootnote}
The relative phase $\Phi$ is well-defined and measurable but is
relevant only in the presence of coupling between the states.
Directly after population transfer it is zero and uniform across
the sample, but with each experimental run it quickly acquires an
arbitrary value, e.g. because of slight irreproducibility in the
trapping potentials.
\end{thebibliography}
\end{document}